# Fabrication of electronically conductive protein-heme nanowires for power harvesting

*Lorenzo Travaglini, Nga T. Lam, Artur Sawicki, Hee-Jeong Cha, Dawei Xu, Adam P. Micolich, Douglas S. Clark, and Dominic J. Glover*[*]

Lorenzo Travaglini, Nga T. Lam, Artur Sawicki, Dominic J. Glover

School of Biotechnology and Biomolecular Sciences, University of New South Wales, Sydney, NSW 2052, Australia.

Corresponding author: Dominic J. Glover (d.glover@unsw.edu.au)

Hee-Jeong Cha, Dawei Xu

Department of Chemical and Biomolecular Engineering, University of California, Berkeley, California 94720, United States.

Dawei Xu

CAS Key Laboratory for Biological Effects of Nanomaterials and Nanosafety, National Center for Nanoscience and Technology, Chinese Academy of Sciences, Beijing 100190, China.

Adam P. Micolich

School of Physics, University of New South Wales, Sydney, NSW 2052, Australia.

Douglas S. Clark

Department of Chemical and Biomolecular Engineering, University of California, Berkeley, California 94720, United States; Molecular Biophysics and Integrated Bioimaging Division, Lawrence Berkeley National Laboratory, Berkeley, California 94720, United States.

**Abstract**

Electronically conductive protein-based materials could enable the creation of bioelectronic components and devices from sustainable and nontoxic materials, while also being well-suited to interface with biological systems, such as living cells, for biosensor applications. In addition, protein materials have other desirable properties such as inherent self-assembly and molecular recognition capabilities. However, as proteins are generally electrical insulators, the ability to render protein assemblies electronically conductive in a tailorable manner could usher in a plethora of useful materials. Here, we present an approach to fabricate electronically conductive protein nanowires by incorporating and aligning heme molecules in proximity along an ultrastable protein filament. The heme-incorporated protein nanowires demonstrated electron transfer over micrometer distances, with conductive atomic force microscopy showing individual nanowires having comparable conductance to naturally occurring bacterial nanowires. The heme-incorporated nanowires were also capable of harvesting energy from ambient humidity when deposited as multilayer films. Exposure of films to humidity produced electrical current, presumably through water molecules ionizing carboxy groups in the protein filament and creating an unbalanced total charge distribution that is enhanced by the presence of heme. A wide variety of other porphyrin molecules exist with varying electrochemical behaviors that could enable the electrical properties of protein assemblies to be tailored, paving the way to structurally- and electrically-defined protein-based bioelectronic devices.

## 1. Introduction

Harnessing proteins to transfer charge carriers may enable the creation of ultra-low power electrical components and devices from sustainable and non-toxic materials.[1] Therefore, the fabrication of bioelectronic devices would benefit from the ability to render protein components electronically conductive in a modular and tailorable manner. In addition, proteins can be genetically engineered to create self-assembling nanostructured scaffolds for the attachment and positioning of functional molecules with specific geometries.[2] Combining charge transfer capabilities with nanostructured protein architectures could enable the creation of protein scaffolds that can interface biological systems such as enzymes or living cells with electronic devices for biosensing, biocatalysis, and biocomputing applications.

Inspiration for the engineering of conductive protein scaffolds can be drawn from bacteria that build highly conductive protein nanowires capable of electron transfer over micrometer-scale distances. A variety of bacterial species in the *Geobacter* and *Shewanella* genera produce protein nanowires to transfer electrons produced during anerobic respiration to extracellular acceptors, such as iron oxides.[3,4] The mechanisms of electron transfer along bacterial nanowires have been associated with electron hopping or tunneling between redox species such as iron-containing heme groups or pi electrons in stacked aromatic amino acids.[5] In the case of nanowires from *Geobacter sulfurreducens*, it was recently demonstrated that electron transfer occurs through sequential electron hopping along a filament composed of hexaheme cytochrome protein subunits.[6] Although bacterial nanowires may be useful conductive materials, these protein complexes are difficult to repurpose, and the mechanisms of their electron transfer are under debate.[7,8]

Inspired by cytochrome-based bacterial nanowires, electrically conductive protein nanowires have been built from the ground up through the attachment and alignment of metalloproteins on non-conductive protein filaments.[9,10] In an example of this strategy, a filament-forming prion protein was recombinantly expressed in fusion with rubredoxin, an iron-sulfur metalloprotein.[9] Films of the resulting rubredoxin nanowires were shown to be electrically conductive and able to mediate electron transfer to incorporated enzymes. Alternatively, protein nanowires have been created using bioconjugation to covalently attach tetra-heme cytochrome c proteins to subunits of an ultra-stable protein filament.[10] The filament scaffold was gamma-prefoldin (γPFD) that originates from *Methanocaldococcus jannaschii*, a deep-sea hyperthermophilic archaeon. Subunits of γPFD assemble through β-sheet domains into helical filaments of up to 3 μm in length, with the remainder of the protein protruding as a coiled-coil domain (**Figure 1a**). Filaments of γPFD are thermostable ($T_m$ of 93°C) and highly engineerable making them ideal scaffolds for functional materials.[11,12] Electrically conductive metalloprotein nanowires were created in a modular approach using the SpyTag-SpyCatcher conjugation system, whereby proteins tagged with the SpyTag peptide spontaneously form amide bonds with proteins fused to the cognate SpyCatcher domain.[13] The cytochrome c3 from *Desulfovibrio vulgaris* was chosen to mediate electron conductance and was produced recombinantly in fusion with a SpyCatcher. Subsequently, nanowires were assembled by conjugating the cytochrome c3-SpyCatcher fusion with γPFD subunits containing a SpyTag domain. Networks of the γPFD-cytochrome c3 nanowires were shown to conduct current at the redox potential of the cytochrome domains.[10] More recently, it was shown the γPFD-cytochrome c3 nanowires were capable of long-range protonic conductivity.[14] The γPFD scaffold is rich in carboxylic acid groups, such as aspartic and glutamic acid groups, which provide a sufficient density of carboxyl groups for the exchange of protons along the nanowire.

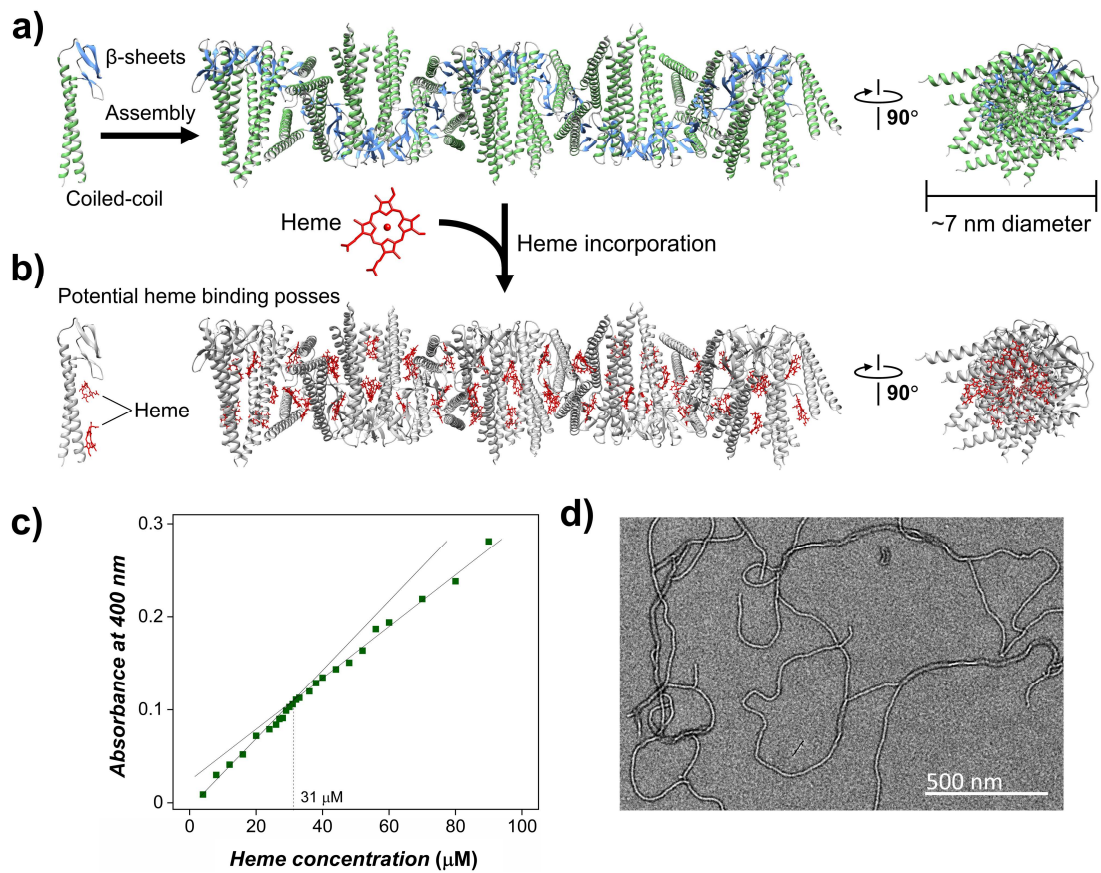

**Figure 1.** Structure of γPFD filaments and incorporation of heme to make conductive nanowires. a) Filament assembly of γPFD through β-sheet domains (PDB entry: 6VY1), and b) proposed binding of heme molecules to the coiled-coil domains to form γPFD-heme nanowires. c) Protein-ligand binding isotherm with 30 μM of γPFD and varying concentration of heme, which indicates a stochiometric ratio of ~ 1 heme per γPFD subunit in filaments. d) TEM image of the γPFD-heme nanowires. Scale bar = 500 nm.

Although the above strategies produced electronically conductive protein nanowires, these strategies required considerable protein engineering and the production of metalloproteins that can be difficult to recombinantly express in host organisms.[15] Rather than attaching metalloproteins to a filament, it should be possible to create conductive protein nanowires by exploiting the natural affinity of some proteins to bind heme or by engineering filamentous proteins to bind heme. Alignment of heme in sufficient density along a protein filament should enable electron transfer between heme groups along the filament. Coiled-coil domains in a variety of proteins have been shown to naturally bind heme, which has been exploited to create synthetic metalloproteins with enzymatic functionalities.[16] The coiled-coil domains of the γPFD filaments that protrude outwards in a helical brush-like arrangement (**Figure 1a**) have been shown to have affinity for a variety of molecules and metals.[11] We hypothesized the coiled-coil domains of γPFD could have affinity to heme, which if aligned in sufficient density and proximity, could mediate electron transfer along the filament.

Herein, we demonstrated that filaments of γPFD can function as scaffolds for heme incorporation to create electronically conductive protein nanowires. Films of γPFD-heme nanowires were highly conductive compared with unmodified γPFD. Individual γPFD-heme nanowires were shown by conductive AFM (c-AFM) to have similar or greater conductance than bacterial nanowires. Multilayer films of bacterial nanowires have been shown previously to be capable of harvesting energy from ambient humidity.[17] Surface carboxy groups in these bacterial nanowires function as a source of exchangeable protons, with water molecules penetrating the film proposed to generate an ionization gradient in the carboxylic groups or a concentration gradient of mobile protons.[17] Creation of a proton gradient and diffusion of protons is proposed to induce a counterbalancing electrical field and current production.[17,18]

As filaments of γPFD have inherent protonic conductivity,[14] we hypothesized that the incorporation of heme would enable or improve energy harvesting from ambient humidity through enhanced charge transfer. Proton transfer in a water solvent occurs through the Grotthuss mechanism,[19] in which protons are transferred from one water molecule to another via formation and breaking of hydrogen bonds. The $Fe^{3+}$ and $Fe^{2+}$ ions in the heme may contribute to proton transfer processes within the nanowire film by providing additional proton hopping sites,[20] thereby increasing the rate of proton transfer, and ultimately enhancing current flow in γPFD-heme films. Additionally, heme molecules containing the charged $Fe^{3+}$ and $Fe^{2+}$ ions have the potential to alter the internal structure of the film,[21,22] which may enable more efficient proton diffusion through the material. The engineerability of the γPFD-heme nanowires also enabled us to examine the impact that heme incorporation has on humidity energy harvesting,[18] which is not possible with natural bacterial nanowires.[17]

The ability to modulate the electronic properties of heme molecules, such as its oxidation state and chemical structure, holds significant potential for tailoring the properties of heme-incorporated conductive biomaterials, for example in applications such as humidity energy harvesting and bioelectronics.[23,24] Furthermore, the ability to render a protein nanostructure electronically conductive through heme incorporation may enable the creation of an innovative class of electroactive materials that possess engineered protein nanostructure while enabling efficient electrical interfacing with both biotic and abiotic materials, paving the way to structurally-defined protein-based bioelectronics.

## 2. Results and Discussion

Prediction of heme binding sites on a γPFD filament was performed *in silico* using DiffDock software,[25] which performs molecular docking using a diffusion generative model over the space of ligand poses. A cryo-EM model of a γPFD filament (PDB entry: 6VY1) consisting of 14 γPFD subunits was used as the protein template and a heme molecule served as the ligand. The molecular docking proposed two main binding locations, with one site positioned between and towards the ends of the coiled-coil domains of four γPFD subunits that protrude outwards from the filament (**Figure 1b**). The second potential binding site is closer to the β-sheet domains, with the heme positioned between the adjacent coiled-coils of two γPFD subunits. Binding of heme to either location on γPFD should produce a sufficient density and proximity of heme along the filament to create electronically conductive protein nanowires.

Encouraged by the molecular docking results, we examined the ability of recombinantly produced γPFD filaments to bind heme. Filaments of γPFD were recombinantly expressed using an *E. coli* host and purified by multimodal chromatography as previously described.[10,26] The filaments were incubated with a molar excess of heme for 16 hours and the filaments purified using either dialysis or size-exclusion chromatography to remove unincorporated heme. Subsequently, the absorption spectrum of the purified γPFD-heme filaments was compared to free heme in solution using UV-Vis spectroscopy. Free heme had a distinct Soret peak at 380 nm associated with the $Fe^{3+}$ ions in the porphyrin molecule (**Figure S1a**). Whereas the absorption peak for the γPFD-heme sample was shifted to 392 nm, indicating the incorporation of heme into the γPFD filament. The binding of heme to proteins such as cytochrome c generally produces a large shift in the Soret peak of the heme in UV-Vis spectra from 380 nm to ~410 nm.[27–29] This notable shift in the absorption peak of a protein-heme complex is generally associated with histidine or cysteine residues in the protein binding the heme group,[30,31] which are the most prominent heme-coordinating residues.[32] The γPFD lacks histidine and cysteine residues, however, it does contain four tyrosine residues that are less common heme-coordinating residues. Hydrophobic interactions with the porphyrin ring system as well as electrostatic interactions and hydrogen bonding via the propionate side chains can contribute to heme binding in proteins.[33] The γPFD is particularly rich in both surface-exposed hydrophobic groups and charged residues due to its role as a molecular chaperone for binding denatured proteins, with these types of residues potentially involved in heme binding by γPFD.

The redox activity of the γPFD-heme nanowires was examined by UV-Vis absorption spectroscopy in the presence or absence of a strong reducing agent. The Soret peak at 392 nm of γPFD-heme in an air-oxidized state was shifted to 422 nm after addition of the reducing agent sodium borohydride (**Figure S1b**). The shift in the Soret peak of γPFD-heme upon reduction is representative of electron transfer to the iron core of the heme, as changes in bond energy within the porphyrin ligand introduce variations in the absorption spectrum.[34–36] Additionally, a broad peak at 560 nm was observed upon nanowire reduction, which might be associated with low-energy transitions in the porphyrin under the reduced state of the heme.[37]

The molar ratio of heme to γPFD subunits in the γPFD-heme nanowires was determined using a ligand-binding isotherm (**Figure 1c**), where the absorption at 400 nm was acquired at different molar ratios of heme mixed with γPFD filaments. The ligand-binding isotherm produced two linear trends at different concentrations of heme relative to γPFD. The initial trend represents the binding of all heme molecules to filaments until saturation of the binding sites occurs, which results in a second trend line due to lower absorbance of free heme at 400 nm (**Figure S1a**). A change in slope for heme binding in ligand-binding isotherms is commonly observed for heme binding to metalloproteins, although the change in slope is generally larger due to these heme-bound proteins having Soret peaks of ~ 410 nm.[27,28] The intersection of the two linear trends for saturation of heme binding to γPFD filaments occurs at 31.8 ± 0.8 μM of heme (average of 3 independent measurements) for 30 μM of γPFD (**Figure 1c**), which suggests a binding stoichiometry of one heme molecule per γPFD subunit in the nanowire. For comparison, free heme in the absence of γPFD produced a single linear trend without any change in slope (**Figure S2a**).

An engineered variant of γPFD with truncated coiled-coils was recombinantly produced to determine the approximate region of heme binding. The variant called γPFD6 has six turns in each alpha helical region that comprises the coiled-coil domain in comparison to the wild-type γPFD that has 12 alpha helical turns.[38] The truncated γPFD6 has been shown to still form filaments as the coiled-coil is physically separate from the β-sheet domains that are responsible for filament assembly. The ligand-binding isotherms for heme incubated with the truncated γPFD6 variant showed no change in the binding slope (**Figure S2c**), which suggests heme was unable to bind to γPFD6 filaments. Therefore, the heme binding site is likely located in the bottom half of the coiled-coil domain. A pair of tyrosine residues (residues 15 and 118) in this region may be involved in the heme binding as tyrosine residues are commonly associated with heme binding in many proteins.[33]

After establishing that heme binds γPFD subunits at a 1:1 molar ratio within filaments, we fabricated all future batches of γPFD-heme nanowires by incubating a 1.2 molar excess of heme with γPFD filaments for 16 hours at 4 °C. Subsequently, γPFD-heme nanowires were dialyzed against phosphate buffered saline (PBS) buffer (150 mM NaCl, 50 mM $NaH_2PO_4$, pH 7.4) to remove unbound heme. The structure and lengths of the γPFD-heme nanowires were compared to unmodified γPFD filaments by imaging the materials using negative-stain TEM (**Figure. 1d**). The binding of heme to γPFD filaments did not change the overall morphology and diameters of the nanowires (**Figure S3a-b-c-d**). The TEM images of the samples revealed a Gaussian distribution of diameters for the γPFD filaments and γPFD-heme nanowires (**Figure S3e-f**). The average diameter of the γPFD filaments was measured to be 9.7 ± 1.7 nm (n = 250 filaments), while the average diameter of the γPFD-heme nanowires was 9.0 ± 1.7 nm (n = 150 nanowires), presented in **Figure S3e**. A t-test was performed and showed no significant difference in diameter between the γPFD filaments and γPFD-heme nanowires. The average length of the γPFD filaments was 1.4 ± 0.6 μm, whereas the average length of the γPFD-heme nanowires was 1.2 ± 0.6 μm (**Figure S3f**). A t-test showed the differences in length was significant, which may be attributed to the additional manipulation required to incorporate heme into filaments.

The morphology of the nanowires in dry conditions was examined using AFM techniques to verify that the negative staining process used in TEM imaging did not affect filament dimensions (**Figure 2a**). The nanowires in the AFM images revealed a consistent height profile of approximately 4 nm, which is indicative of the nanowire diameter when dry. Solid state electrical measurements were performed on dry films of γPFD filaments or γPFD-heme nanowires deposited on interdigitated titanium/gold microelectrode (IDME) devices (**Figure 2b**). The IDMEs were fabricated on $Si/SO_2$ substrate using lithography techniques and featured 50 electrodes spaced 3 μm apart, each with a length of 0.5 mm, resulting in total electrode lengths of 2.5 cm (**Figure S4a**). Protein samples at a concentration of 60 μM were drop casted onto IDME devices and dried under atmospheric conditions overnight before performing electrochemical measurements (a scheme of the device fabrication process is presented in **Figure S4b**). The deposition of a homogeneous nanowire film between electrodes was examined by AFM, which showed an interconnected network of evenly distributed γPFD-heme nanowires over the whole area (**Figure 2b**).

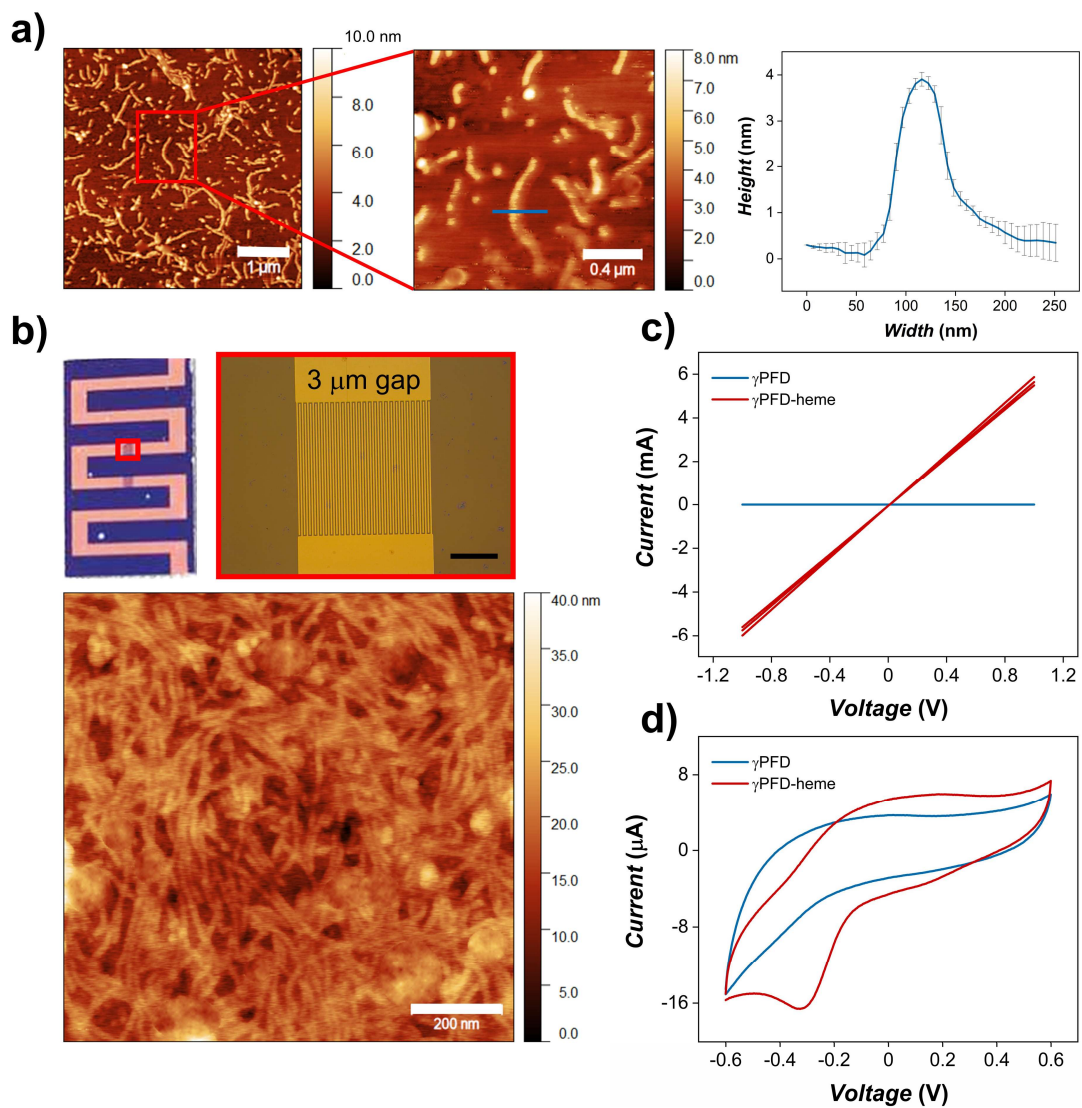

**Figure 2.** Characterization and electrochemical behavior of γPFD-heme nanowire networks on interdigitated microelectrodes. a) Morphology of γPFD-heme nanowires imaged by AFM (scale bars of 1 μm and 0.4 μm, respectively) and measured height and width profile of a single γPFD-heme nanowire. b) Optical image of the patterned interdigitated microelectrode and AFM images showing the structure of γPFD-heme nanowire films deposited between electrodes. Scale bar = 200 μm. c) *I-V* curves of the unmodified γPFD filaments (blue curve) and γPFD-nanowires (red curve) measured on microelectrodes by sweeping the voltage between −1 and +1 V. d) Cyclic voltammogram of unmodified γPFD filaments (blue curve) and γPFD-heme nanowires (red curve) measured by sweeping the potential between −0.6 V and +0.6 V and with a scan rate of 50 mV/s.

The *I-V* characteristics of the dry films were measured by sweeping the voltage between the two IDME electrodes between −1 V and +1 V and recording the output current. The γPFD-heme nanowire films exhibited a linear *I-V* behavior, with a calculated resistance of 175 ± 5 Ω and conductance of 5.7 ± 0.2 mS (**Figure 2c**). Films of unmodified γPFD filaments were orders of magnitude less conductive, with a resistance of approximately 440 kΩ and conductance of 2.3 µS (**Figure 2c**). Thus, the enhanced conductivity of the γPFD-heme nanowires can be attributed to the presence of redox active heme centers that enable efficient electron transfer along the nanowires. The conductive nature of the nanowires outperform natural bacterial nanofibers such *G. sulfurreducens* pili films with a measured resistance of 470 Ω,[39] and have similar conductivity as engineered aromatic-rich peptide nanofibers with a resistance of 190 Ω.[40]

Electron transfer through the films on the IDME devices should occur not only along individual nanowires but also between nanowires, as the nanowires are shorter than the 3 µm gap of the gold IDME electrodes (**Figure 2b**). To investigate this phenomenon further, we conducted electrical measurements with lower concentrations of deposited γPFD-heme nanowires of 4 and 2 µM on IDME devices. The purpose of these measurements was to demonstrate that a network of interconnected nanowires is required to bridge the electrodes to observe a non-negligible current. Nanowires were deposited at 4 µM or 2 µM concentrations on the IDME devices, in comparison to the 60 µM concentration that was used to create films. Subsequently, the *I-V* characteristics of the protein-coated decorated IDMEs were recorded, and the film morphology imaged using AFM to examine the structure of the nanowire network.

Samples prepared with 4 µM of nanowires had interconnected individual nanowires that formed contiguous paths between the electrodes of the IDME devices (**Figure S5 a-b**). In contrast, nanowires deposited at 2 µM on IDMEs resulted in isolated individual nanowires that do not connect between the electrodes (**Figure S5 c-d**). The contiguous paths of nanowires at 4 µM were shown to enable current flow between the electrodes (**Figure S5e**), whereas a lack of nanowire connection between electrodes resulted in no conductivity when a potential was applied to the IDME (**Figure S5f**). We can conclude that long-range conductivity over a length of several µm requires an interconnected network of conductive nanowires to facilitate inter-nanowire charge conduction to bridge and connect the two gold electrodes. Previous studies have demonstrated that parallel alignment of conductive nanowires can enhance charge transport.[41] In a similar manner, the conductivity of films composed of randomly orientated γPFD-heme nanowires might be further enhanced if the orientation of nanowires can be ordered and aligned between electrodes.

Potentiometric measurements were performed to examine the oxidation and reduction potential of the γPFD-heme films. A three-electrode setup was used for the measurements that consisted of a glassy carbon working electrode, a platinum counter electrode, and an Ag/AgCl reference electrode. The redox properties of the film deposited on the working electrode were probed with cyclic voltammetry (CV) in PBS as the electrolyte (**Figure 2d**). Applying voltage sweeps from −0.6 to +0.6 V produced a prominent peak at −0.3 V that is associated with the reduction of $Fe^{3+}$ to $Fe^{2+}$ present in the heme molecule. A slight peak was also observed at +0.3 V which is associated with the oxidation of $Fe^{2+}$ to $Fe^{3+}$. The lower intensity of the oxidation peak is presumably caused by the measurements being performed in ambient atmosphere with the incorporated heme in the γPFD-heme complex being partially oxidized.

The samples were also examined by CV under anaerobic conditions and the voltammogram presented in **Figure S6** shows well-defined peaks associated with the anodic and the cathodic voltametric peaks of both $Fe^{3+} \rightarrow Fe^{2+}$ and $Fe^{2+} \rightarrow Fe^{+}$ redox processes of the embedded heme.[42] Previous studies investigating heme-containing myoglobin films have reported the gradual loss of heme over time when exposed to an electrolyte environment,[43] however no decrease in the magnitude of redox peaks was observed for γPFD-heme nanowires after repeated cycles, which suggests there was no heme loss occurring during CV.

Electrochemical impedance spectroscopy (EIS) was performed to probe the capacitive and resistive nature of the protein materials with the same three electrode setup used for CV. Overall, the EIS analysis revealed resistive behavior of the γPFD-heme nanowires films in contrast with the more capacitive characteristics of the unmodified γPFD films, as shown in **Figure 3**. The Nyquist plot shown in **Figure 3a** probes the resistance of the nanowires to charge transfer and the capacitance-impedance regimes. The γPFD-heme nanowire films had a smaller semicircle for high frequencies (**Figure S7**), which confirms a lower resistance to charge transfer when compared to γPFD films. Additionally, unmodified γPFD films manifested a 45° angle trend at lower frequencies that suggests a Warburg response, which is a mixed capacitance-impedance behavior dominated by the ionic diffusion between the electrolyte and the film, typical of a more insulating material. Although the γPFD films had a Z' of about 5 kΩ which is comparable with previously reported conductive protein biofilms with similar experimental conditions,[44] the γPFD-heme films exhibit significantly lower magnitude Z' values, which demonstrates a more conductive material in comparison with previously engineered protein based biopolymers.[41,45] The Bode plot shown in **Figure 3b** revealed that the magnitude of impedance (solid lines) at low frequencies is lower for γPFD-heme films in comparison with the γPFD films. In addition, the magnitude of the phase shift of γPFD-heme films (red dotted lines) showed a decrease at intermediate frequencies (1-10$^3$ Hz), which indicates a shift from capacitive behavior to a more resistive character for γPFD-heme nanowire films. This tendency is confirmed by estimating the crossover frequency, which is the frequency at which the magnitude and phase of the impedance intersect. The crossover frequency at a higher frequency for the unmodified γPFD filaments films indicates the system is dominated by capacitive effects. The impedance of a capacitor decreases as frequency increases, so at higher frequencies, the capacitive effects will dominate, and the magnitude of impedance will be low. Conversely, the crossover frequency for γPFD-heme films is at a lower frequency and may indicate that the system is dominated by more inductive effects.

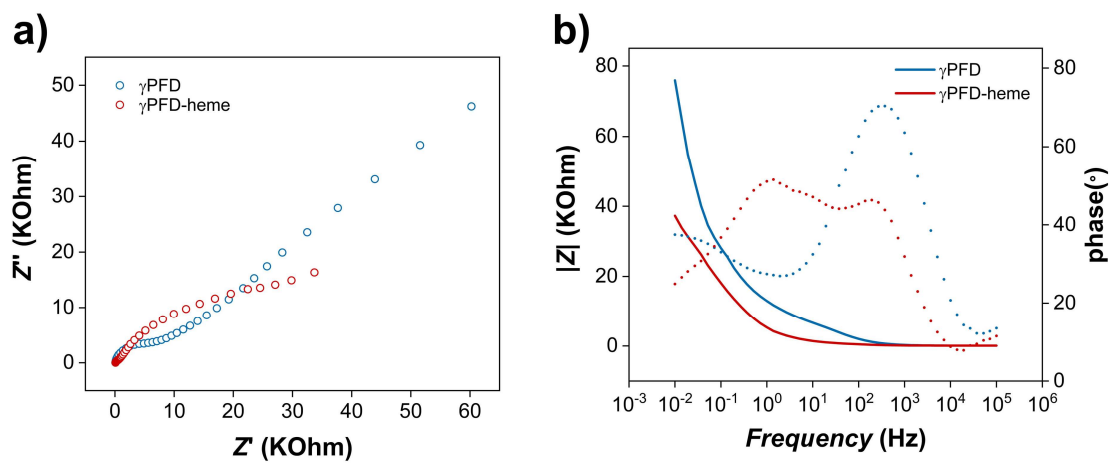

**Figure 3.** Electrochemical impedance analysis of films composed of unmodified γPFD filaments or γPFD-heme nanowires. a) Nyquist plot and b) Bode plot of unmodified γPFD filaments and γPFD-heme nanowire films marked in blue and red, respectively. In the Bode plot, solid lines refer to the absolute value of the impedance and dotted lines refer to the phase shift.

Conductive atomic force microscopy (c-AFM) was used to probe the conductive properties of individual γPFD-heme nanowires. In the c-AFM setup shown in **Figure 4a**, a potential difference was applied across a nanowire in contact with a surface electrode at one end and the AFM tip. Varying tip position along the nanowire relative to the electrode unveils resistive behavior; increased distances raise resistance and lower current, shorter distances decrease resistance and heighten current, enabling systematic assessment of nanowire conductivity.

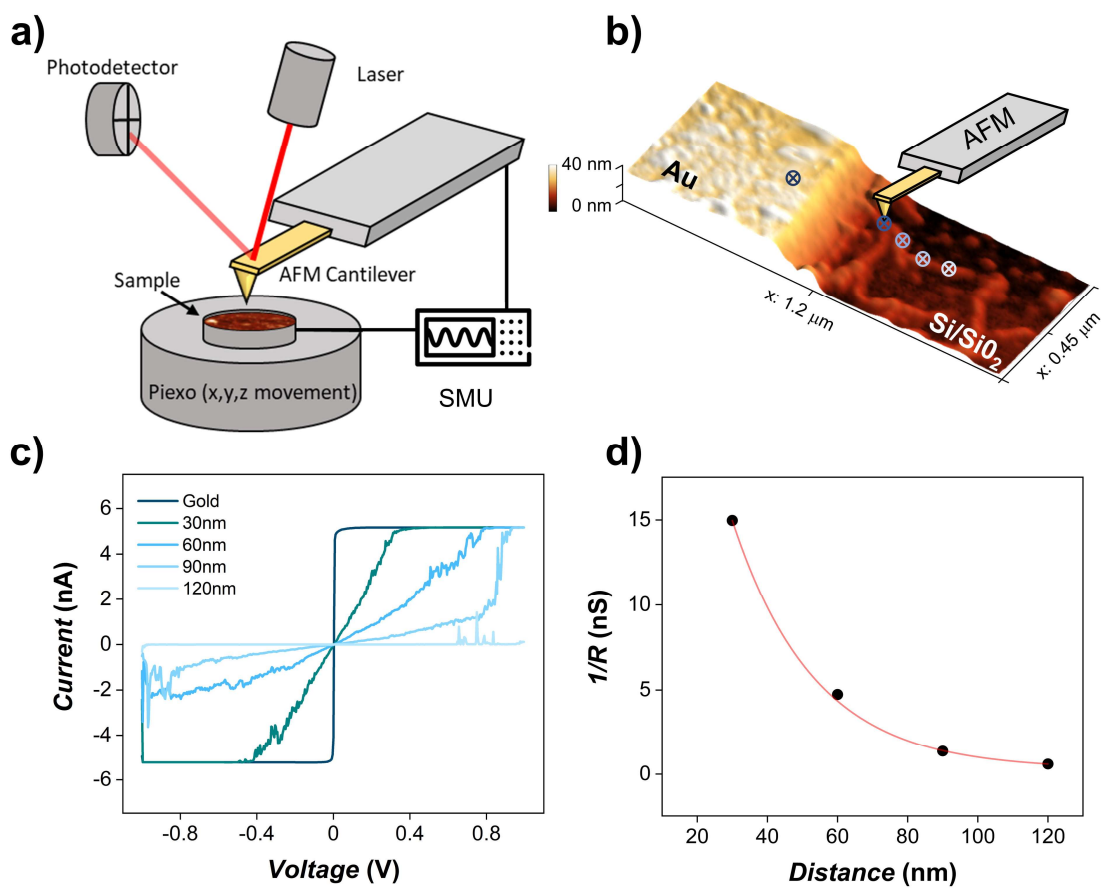

**Figure 4.** Electrical characterization of individual γPFD-heme nanowires using conductive AFM (c-AFM). a) Illustration of the c-AFM setup for the electrical characterization of γPFD filaments and γPFD-heme nanowires. b) AFM image and scheme of *I-V* measurements performed along the 3D morphology of a γPFD-heme nanowire that is in contact at one end with a gold electrode. The positions along the nanowire and gold electrode where *I-V* measurements were performed with the AFM tip are shown by crossed circles. c) *I-V* curves measured with the conductive tip positioned on the gold electrode or at different distances along the nanowire from the electrode. d) Conductivity of a γPFD-heme nanowire versus the distance from the gold electrode.

A 2 μM solution of γPFD-heme was deposited on devices consisting of gold IDME on Si/SiO$_2$ substrates. The nanowires were allowed to settle on the device surface for 5 minutes, rinsed with deionized water to remove residual salt, and air-dried for 30 minutes. Subsequently, AFM was used to map the surface morphology of the IDME to locate individual nanowires in contact with the gold electrode while also extending across the insulating Si/SiO$_2$ substrate (**Figure 4b**). The AFM tip was positioned in contact with a nanowire at varying distances (0 to 120 nm) from the electrode as shown in **Figure 4b**, and a potential sweep applied between −1 V and +1 V to measure the current response at each point of the nanowire. The current between the gold electrode (connected to the AFM stage) and the AFM tip at different distances along the nanowire exhibited linear behavior with slopes that decrease with an increase in the distance of the tip from the electrode (**Figure 4c**). The nanowire resistance was determined from the slopes of the *I-V* curves and its conductance as a function of nanowire length was calculated and is presented in **Figure 4d**. The average conductivity of three individual nanowires was determined by c-AFM to be $0.3 \pm 0.1$ S·cm$^{-1}$ (**Figure S8**). The results confirmed that single γPFD-heme nanowires showed ohmic behavior with a conductance decrease with the nanowire length. The conductivity of individual γPFD-heme nanowires is comparable with previously reported peptide nanowires of $1.12 \pm 0.77$ S·cm$^{-1}$[40] and significantly higher when compared with *Geobacter pili* at $0.051 \pm 0.019$ S·cm$^{-1}$[46]

The *I-V* response of individual unmodified γPFD filaments was measured for comparison with the nanowire behavior using the same experimental setup. The γPFD filaments showed negligible current at all distances measured along the filaments (refer to **Figure S9**), therefore their conductance was not calculated. Overall, the electrical characteristics of individual nanowires and filaments aligns with the *I-V* characteristics of the macroscopic films of γPFD-heme nanowires and unmodified γPFD filaments. This agreement in conductivity between the individual nanowires and the macroscopic films highlights the consistency in electrical behavior across different length scales within the γPFD-based materials.

Having confirmed that the engineered γPFD-heme nanowires can serve as an electroactive and conductive material, we demonstrated the nanowires can function as an energy harvesting material that generates electrical energy from moisture in the environment. Films of bacterial nanowires have been shown previously to harvest energy from ambient humidity through the formation of a moisture gradient within the film, which creates unbalanced charge and produces an electric output.[17,18,47] Our approach takes advantage of the γPFD being rich in carboxylic groups, which can serve as a source of exchangeable protons.[14] We hypothesize that a moisture gradient is generated within the film as water molecules penetrate the material. Water molecules ionize the COOH groups of γPFD, which should generate a gradient of COO$^-$ immobile ions and H$^+$ mobile ions that diffuse through the material, as represented in **Figure 5a**. This diffusion process should be facilitated from the γPFD filament backbone of the nanowires possessing a hole-like conduction, as we recently demonstrated.[14] An unbalanced total charge distribution should be produced in the hydrated film, thereby causing an electric field and a consecutive potential difference between the surface and the bottom of the film. Placement of electrodes at these locations on the film produces an open circuit voltage $V_O$ that can be measured, and a short circuit current $I_{SC}$ is established by connecting the two electrodes.

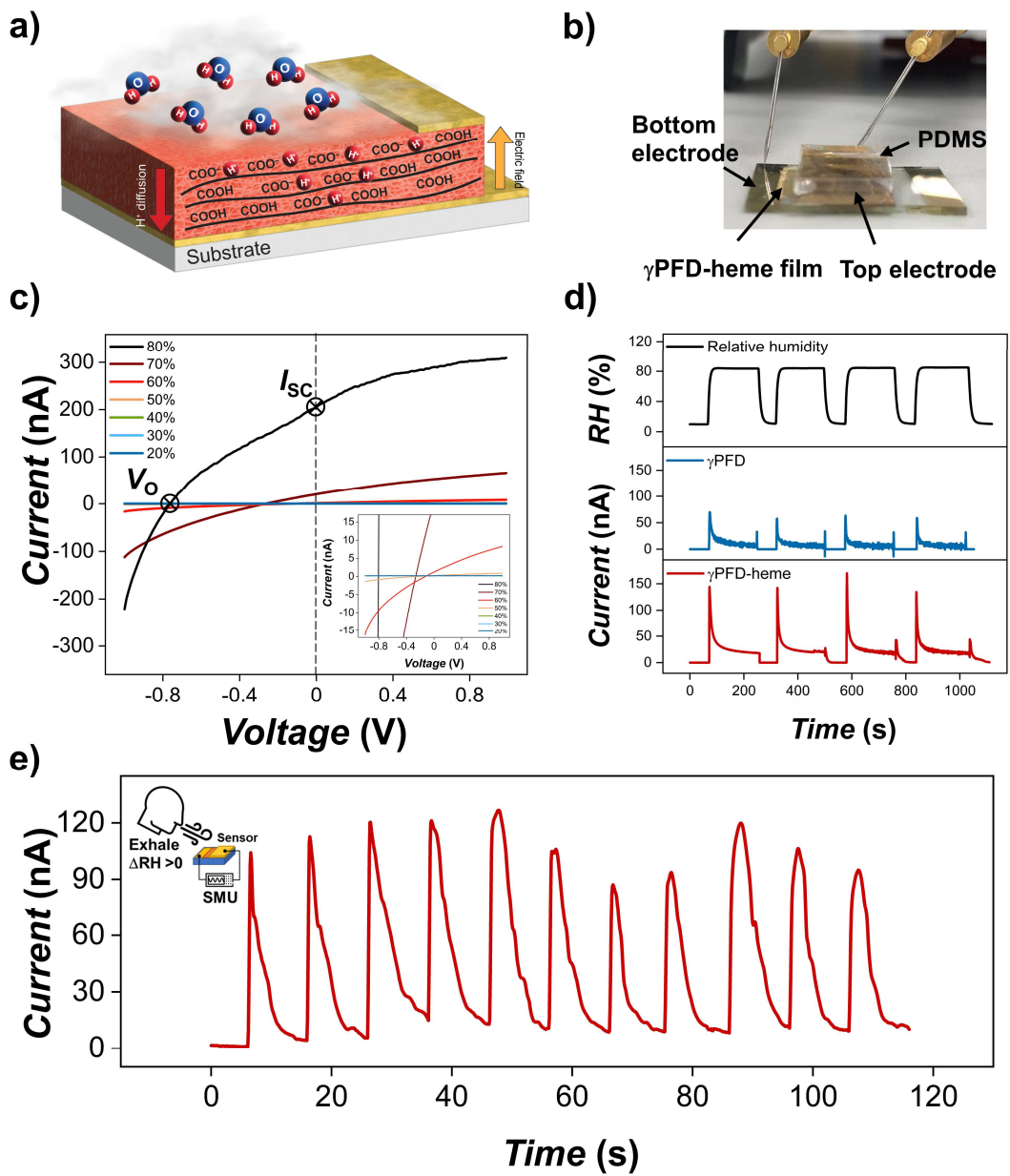

**Figure 5.** Design and real-time performance of a γPFD-heme nanowire-based humidity harvesting device. a) Diagram of the humidity harvesting device structure and proposed electronic/protonic transport mechanisms. b) Photograph of a fabricated humidity harvesting device that contains a γPFD-heme nanowire film. c) *I-V* curves of the γPFD-heme nanowire-based power generator at relative humidity (RH) ranging from 20% to 80%. d) Continuous recording of a generated current over time at RH varying from 10% to 80%. e) Real-time response of a γPFD-heme to human breath. Each spike in the current production corresponds to breath applied to the sensor.

Protein nanowire-based devices were fabricated with a bottom gold electrode deposited on a glass substrate, and a solution of heme nanowires drop casted in a confined PDMS well to control the film dimensions (0.5 cm x 2 cm) and dried in an oven at 60 °C for 1 hour to obtain a film thickness of ~ 7 μm. A gold coated PDMS was positioned onto the film serving as a top electrode as shown in **Figure 5b**. In addition, devices were constructed that contained unmodified γPFD filament films to compare their current generation against the heme-containing nanowire films. An atmospheric control chamber was built to enable the electrical properties of the device to be tested with varying 20% to 80% relative humidity (RH) (**Figure S10**). The vacuum chamber was designed with a water bubbler to regulate the RH inside the chamber, and a Sensirion SHT2x capacitive humidity sensor was installed in the sample chamber to continuously monitor the RH.

The electrical properties of the γPFD-heme devices at varying RH were examined by sweeping the potential between the top and bottom electrode and recording *I-V* curves (**Figure 5c**). The γPFD-heme devices had mostly linear *I-V* characteristics for RH ranging from 20% to 40% (**Figure 5c** inset) with no obvious changes in current magnitude, whereas a deviation from the linear behavior was observed for RH of 50% to 80% with a pronounced increase in the current magnitude. This trend has been previously shown in naturally occurring bacterial protein nanowire based biofilms.[3] Furthermore, the *I-V* curves did not pass through the origin (see **Figure 5c** inset), which enabled an estimation of an open circuit voltage ($V_O$) and short circuit current $I_{SC}$. The curve shift from the origin was directly dependent on the chamber RH, with 80 % RH producing an $I_{SC}$ of +205.1 nA and $V_O = -0.76$ V, as marked in **Figure 5c**. A power density of 223 μW cm$^{-3}$ was calculated when taking the device geometry into consideration. The power density of the γPFD-heme device was orders of magnitude greater than many carbon-based devices that produce current from humidity,[48,49] and comparable to natural protein-based nanofiber devices.[17,50,51] This peculiar behavior has been also shown for natural bacterial protein nanowires harvested from *Geobacter sulfurreducens* that were used to fabricate a power generator device from ambient humidity.[17]

The current generation of devices composed of γPFD-heme nanowires or unmodified PFD filaments were compared over time with changes in RH from 10% to 80%. Without any applied bias, the γPFD-heme devices produced significantly more current than devices composed of γPFD filament and superior signal to noise ratios for consecutive RH changes (**Figure 5d**). Devices containing unmodified γPFD filaments showed a slight increase in current from 0.020

± 0.008 nA at 10% RH to 6.570 ± 2.240 nA at 80% RH after 3 minutes, which gives a wet:dry current ratio of 328. By comparison, the γPFD-heme nanowire-based films devices produced 0.017 ± 0.007 nA at 10% RH increasing to 21.099 ± 1.310 nA at 80% RH giving a wet:dry ratio of 1241, which is 3.8 times higher than the γPFD devices that do not contain heme. The response times of the unmodified γPFD filaments and nanowire-based devices to produce maximum current were measured as a single step from a RH of 10% to 80% (**Figure S11**). The response time was determined to be ~ 3 s for the γPFD filament-based device compared to the more rapid response time ~ 1.2 s for the γPFD-heme device.

The increase of RH to 80% presumably produces an initial surge in the electric current due to the accumulation of positively charged protons ($H^+$) at the electrode/material interface. Diffusion leads to the accumulation of positive charge at the lower surface and a presence of negative anionic charge at the upper surface, which gives rise to an electric field that opposes the charge diffusion. The nature of charge dynamics in materials that harvest energy from humidity remains a topic of ongoing discussion. While the precise ionization or concentration of $H^+$ ions is influenced by molecular dynamics necessitating additional exploration, a logical supposition is that ionization could correlate with the presence of water molecules or the moisture absorption ratio within the nanowire film. As a result, it is reasonable to anticipate that the open-circuit voltage ($V_O$) would correspondingly vary with the moisture disparity between the upper and lower interface, which was observed to occur with γPFD-heme films at varying RH (refer to **Figure 5C**). We conclude that the incorporation of heme into γPFD filaments plays a pivotal role in facilitating charge transfer, resulting in improved humidity-driven current generation and reduced time required to reach maximum current production. Moreover, in addition to the enhanced proton transport facilitated by the Grotthuss mechanism, the presence of heme centers within the filaments may further enhance proton conductivity, thereby contributing to the efficiency of humidity energy harvesting through various supplementary mechanisms. Water molecules absorbed into the γPFD-heme nanowire matrix should interact with the heme molecules and potentially facilitate proton transfer in conjunction with the Grotthuss mechanism. In this arrangement, the heme acts as a hopping site for protons, increasing the proton transfer rate and consequently the conductivity of the bulk γPFD-heme film.

It has been shown in ionic systems that charged redox active sites can facilitate the Grotthuss mechanism by creating a shortcut for charge transport.[52] Dissociation of water molecules to produce an hydroxide anion and a proton can be facilitated by the presence of highly charged ion clusters, such as $Mg^{2+}$-water complexes, with the charged ion influencing the hydrogen bond network.[53] Recent works has proposed that similar mechanisms for proton conduction can occur in natural conductive proteins[41] and engineered proteins,[14] as well as Nafion, a tetrafluoroethylene-based fluoropolymer-copolymer known for its high proton conductivity.[54,55] The morphology of the γPFD-heme film may also influence its proton conductivity and enable greater proton exchange in the bulk material compared with unmodified γPFD filaments. The charged $Fe^{3+}$ and $Fe^{2+}$ ions of the heme groups (depending on the oxidation state) may change the internal structure of the γPFD filament backbone and thereby enable more efficient proton diffusion. For example, the heme groups may promote the interaction of individual coiled coil domains that are known to have flexibility towards their termini.[10] Structural changes have been shown to occur for Nafion at high levels of hydration that cause sulfonic acid groups and water molecules to form structured clusters or domains. These ordered structures create a network of interconnected proton-conducting channels that extend throughout the polymer matrix.[56]

The ability of the γPFD-heme devices to produce electrical current from humidity is potentially amenable for applications such as breath rate monitoring where the humidity involved in the breath is close to a RH of 100%. We therefore examined the capability of the γPFD-heme device to measure the amount of current produced from human exhalation. In our experimental setup, we connected the nanowire device to a source measure unit, applying a constant potential of +0.3 V between the top and bottom electrodes. We specifically chose this voltage to ensure a significant variation in current when under the influence of humidity from breath. Changes in current were recorded as participants exhaled perpendicularly to the device surface, at a distance of approximately 5 cm away. As shown in **Figure 5e**, the γPFD-heme devices functioned as a breath rate sensor, with reliable current output reaching ~ 80 to 100 nA in response to an expiration event (one event every 8-10 s to enable a full recovery of the device to the initial current). The asymmetric shape of the current peak suggests a faster rate of adsorption and slower desorption of water molecules at high RH. For practical applications, a device that functions as a breath sensor does not need to completely absorb and desorb water molecules from each breath; the device only needs to respond momentarily to each expiration event. The response of the γPFD-heme humidity sensor was also recorded at short 3-4 s breathing rates (**Figure S12**). The output of the device does not recover to the initial current value, but evident variation of the output current was recorded during each expiration event, which demonstrates potential applications for monitoring of breath rate.

Changes in the device geometry such as reduction of the electrode separation and size of the active area may improve the response rate. Alternatively, increasing the device surface area may increase the current production. By connecting multiple devices in series, the generated currents add up, resulting in an increased overall current. This approach is commonly used in various electrical systems to boost output.[17] Regardless, we demonstrated the potential application of this novel material based on engineerable γPFD-heme protein nanowires as electroactive components in bioelectronic devices.

## 3. Conclusion

Incorporation of heme into protein filaments enabled the creation of protein nanowires that were electronically conductive over long distances and capable of harvesting energy from ambient humidity. The fabrication of protein-based electronic and protonic materials through heme incorporation should enable the engineering of innovative electroactive materials capable of electrically interfacing with biological systems, paving the way for a new class of protein-based bioelectronics. Furthermore, the rendering of proteins conductive through heme incorporation should be applicable to other proteins beyond γPFD filaments to create a variety of conductive protein nanostructured assemblies. The electrical properties of protein assemblies that align heme molecules should also be tunable as the charge transfer behavior of heme can be modulated by varying its oxidation state, chemical structure, or the surrounding environment.[57,58] This may enable tuning the sensitivity, response time, or output power of protein-based energy harvesting devices. For example, heme-based materials have been reported to exhibit tunable electrical properties that can be modulated by changing the oxidation state of heme, enabling precise control over the material's electrical characteristics.[23,24] This tunability will be advantageous for optimizing the electrical properties of the protein-based bioelectronic materials for applications such as for humidity energy harvesting.

In the future, it would be interesting to explore the potential of other porphyrin molecules with structures similar to heme but possessing varying electronic properties. Such molecules may exhibit an affinity for protein nanostructures like the γPFD filament. For instance, chlorophyll b could be investigated for light harvesting purposes. However, further research is needed to fully understand and harness the potential of these molecules. Ultimately, by precisely aligning porphyrin molecules within protein nanostructures, we can tailor their electronic properties and unlock the creation of functional protein-based bioelectronics. This interdisciplinary field holds immense promise for developing cutting-edge technologies that bridge the gap between biological systems and electronic devices.

## 4. Experimental Section/Methods

*Computational predication of heme docking:* The binding of heme to subunits of γPFD in filaments was predicted *in silico* using Diffdock,[25] with the PDB entry 6VY1 used as the protein template and a heme molecule (Pubchem ID: 53629486) serving as the ligand. The Diffdock program was performed with 20 inference steps, 40 samples per protein complex, and a batch size of 10. The ligand poses were analyzed using DiffDock and Smina (a fork of AutoDock Vina software),[59] and the top ligand poses based on confidence scores and predicted binding affinities were visualized with UCSF ChimeraX.[60]

*Recombinant protein production:* Filaments of γPFD were expressed in BL21 T7 Express *E. coli* as described previously.[10,26] Briefly, cultures of BL21 containing an expression plasmid were grown at 37 °C in Luria Broth supplemented with 100 μg mL$^{-1}$ ampicillin. Expression of γPFD protein was induced when the culture reached an optical density of 0.6 at A600 with the addition of isopropyl β-d-1-thiogalactopyranoside to a final concentration of 0.4 μM. Harvested cells were suspended in 150 mM NaCl, 50 mM NaH$_2$PO$_4$, pH 7.4, and lysed via sonication (Branson 250, pulse time of 2 s on and 2 s off at a 50% amplitude). The lysate was clarified by centrifugation (25,000 *g* for 30 min) before purification by multimodal chromatography using Capto Core 700 resin (Cytiva).[26] Fractions containing the desired pure protein were pooled and concentrated using 100 kDa Amicon Ultra-15 centrifugal filters, and analyzed via SDS-PAGE with gels stained using Invitrogen SimplyBlue stain. Purified protein was aliquoted into tubes and stored at 4 °C. The truncated variant of γPFD described previously[38] was expressed and purified in the same manner as the wild-type γPFD.

*γPFD-heme fabrication and assembly:* The γPFD-heme nanowires were fabricated by combining a 1.2 molar excess of heme with γPFD filaments and incubation for 16 h at 4 °C. Unincorporated heme was removed via dialysis using Thermo Scientific™ Slide-A-Lyzer™ MINI Dialysis Devices, 10 kDa molecular weight cutoff and dialyzing against 150 mM NaCl, 50 mM NaH$_2$PO$_4$, pH 7.4. Alternatively, free heme was removed by size-exclusion chromatography with a HiTrap Sephadex G-25 resin. Subsequently, the purified nanowires were stored at 4 °C.

*Transmission Electron Microscopy:* Protein filaments and nanowires were diluted to a concentration of 70 µg/mL in PBS and applied onto glow-discharged carbon type-B, 200 mesh copper grids (Ted Pella Inc.). The grids were incubated with the protein for 2 min at room temperature, followed by a wash with 1 mL of milli-Q water. Excess water was removed by soaking with filter paper (Whatman), and the grids were stained with 0.5 mL of a 2% aqueous solution of uranyl acetate (BHD Chemicals) for 2 min. Any excess stain was absorbed with filter paper (Whatman), and the grids were air-dried at room temperature before visualization using a Tecnai G2 20 transmission electron microscope. The obtained TEM images were analyzed using ImageJ software (U.S. National Institute of Health) to measure the length of the filaments and nanowires.

*Quantification of heme incorporation into γPFD filaments:* A series samples were prepared containing 1 mL of γPFD filaments at 30 µM. Varying concentrations (4 µM to 150 µM) of heme dissolved in DMSO were added to γPFD filaments and incubated overnight at 4 °C. Subsequently, the UV-Vis absorption spectra of the samples were obtained using either a Lambda 950 UV-Vis spectrophotometer or a NanoDrop One.

*Microelectrode fabrication:* Silicon n-doped wafers (4") were purchased and a $SiO_2$ layer thermally grown on top of the wafer to a thickness of 250 nm using well-established methods. The wafers were cut into 10 mm x 6 mm chips and a photoresist was hard-baked on the polished side. The photoresist was patterned using a DMO ML3 Pro direct writer and subsequently developed with RU2020 photoresist developer. A 5 nm titanium layer was deposited on the chip, followed by 30 nm layer of gold using a Lesker Electron-Beam Evaporator PVD75, and subsequently, the photoresist was removed in acetone to obtain interdigitated 5 µm electrodes with 3 µm gap distance between electrodes.

*Solid-state I-V measurements of nanowires: I-V* measurements were performed using a PGSTAT204 potentiostat using a 2-electrode setup. Nanowire samples (volume 10 µL and concentration 1 mg/mL) were deposited on the fabricated interdigitated electrode for 5 min and rinsed with deionized water to remove the salt content. Potential sweeps were applied from −1 V to +1 V and the current measured as a function of the potential applied with a scan rate of 50 mV/s.

*Cyclic voltammetry (CV) and electrochemical impedance spectroscopy (EIS):* CV and EIS were performed using an electrochemical cell consisting of a platinum wire, an Ag/AgCl electrode in 0.1 M $AgNO_3$, and a glassy carbon as counter, reference, and working electrodes, respectively. Nanowire or filament films were deposited on the glassy carbon electrodes by drying 10 μL of a 1 mg/mL protein solution overnight at 4 °C. The electrode was immersed in electrolyte solution (150 mM NaCl, 50 mM $NaH_2PO_4$, pH 7.4). The CV was performed by applying potential sweeps between −0.6 V to +0.6 V with a scan rate of 50 mV/s using a PGSTAT204 potentiostat. EIS were performed applying a DC perturbation centered at −0.3 V with amplitude of 0.01 V for a frequency range spanning from 0.01 Hz to 100 kHz using a PGSTAT204 potentiostat.

*Conductive atomic force microscopy (c-AFM):* Imaging of nanowire morphology and nanowire conductive measurements were performed using a Bruker ICON AFM. Coated tips (FMV-PT, Bruker AFM probes) with a nominal spring constant of 2.8 N/m were employed. The AFM was operated in peak force "Tuna" mode and *I-V* measurements along the nanowires were measured using the point and shoot function, which involved sweeping the voltage between −1 V to +1 V and recording the current as a function of the potential applied. A ramp deflection setpoint of 0.5 V, "Current sens" parameter of 1 nA/V and ramp rate of 0.5 Hz were employed to perform the local *I-V*.

*Humidity sensor fabrication:* Using a PDMS well, a 200 μL solution containing nanowires at a concentration of 1 mg/mL was deposited onto a 30 nm thick gold electrode on glass substrate. The deposition was followed by overnight drying under ambient conditions, resulting in the formation of a multilayer film comprised of nanowires. A layer of gold was deposited onto a PDMS substrate, which had dimensions of 1 cm by 2 cm. The coated PDMS was engineered to function as a top electrode that contacts the surface of the nanowire film.

*Humidity sensor characterization:* An atmospheric control chamber was built to enable electrical measurements as a function of hydration/relative humidity (RH) (**Figure S10**). The chamber was built with an inlet connected to a deionized water reservoir to control RH inside the chamber using a valve and an outlet to exhaust gases. A Sensirion SHT2x capacitive humidity sensor was installed inside the chamber for continuous monitoring of the RH.

**Supporting Information**

Supporting Information is included below.

**Acknowledgements**

The authors acknowledge the facilities and scientific assistance of the Mark Wainwright Analytical Centre, including the UNSW Recombinant Products Facility for assistance in the production and purification of proteins; the Electron Microscopy Unit to perform transmission electron microscopy; the Katharina Gaus Light Microscopy Facility for atomic force microscopy and NSW node of the Australian National Fabrication Facility (ANFF) for electrodes fabrication. We acknowledge Professor Kourosh Kalantar Zadeh for providing the setup for conductive atomic force measurements, Professor Damia Mawad for providing the setup for the UV-Vis measurements, Dr. Ky V. Nguyen and Dr. Jan G. Gluschke for assistance with the humidity chamber set-up. D.J.G. is supported by the Office of Naval Research Global (N62909-21-1-2019). D.J.G. and D.S.C are supported by the Air Force Office of Scientific Research (FA9550-20-1-0389).

**Author Contributions**

L.T. designed experimental procedures, characterized the protein materials with AFM, performed electrochemical measurements, and wrote the manuscript with input from all authors. N.T.L. and A.S. produced the recombinant proteins and assisted with nanowire fabrication. N.T.L performed TEM imaging of the nanowires. H.J.C. and D.X. designed mutant recombinant proteins. D.X. designed the nanowires and performed initial heme binding experiments. A.P.M. assisted with the electrical and electrochemical measurements. D.S.C. and D.J.G. designed and supervised the experimental procedures and edited the manuscript.

*Lorenzo Travaglini, Nga T. Lam, Artur Sawicki, Hee-Jeong Cha, Dawei Xu, Adam P. Micolich, Douglas S. Clark, and Dominic J. Glover[*]*

Lorenzo Travaglini, Nga T. Lam, Artur Sawicki, Dominic J. Glover

School of Biotechnology and Biomolecular Sciences, University of New South Wales, Sydney, NSW 2052, Australia.

E-mail: d.glover@unsw.edu.au

Hee-Jeong Cha, Dawei Xu

Department of Chemical and Biomolecular Engineering, University of California, Berkeley, California 94720, United States.

Dawei Xu

CAS Key Laboratory for Biological Effects of Nanomaterials and Nanosafety, National Center for Nanoscience and Technology, Chinese Academy of Sciences, Beijing 100190, China.

Adam P. Micolich

School of Physics, University of New South Wales, Sydney, NSW 2052, Australia.

Douglas S. Clark

Department of Chemical and Biomolecular Engineering, University of California, Berkeley, California 94720, United States; Molecular Biophysics and Integrated Bioimaging Division, Lawrence Berkeley National Laboratory, Berkeley, California 94720, United States.

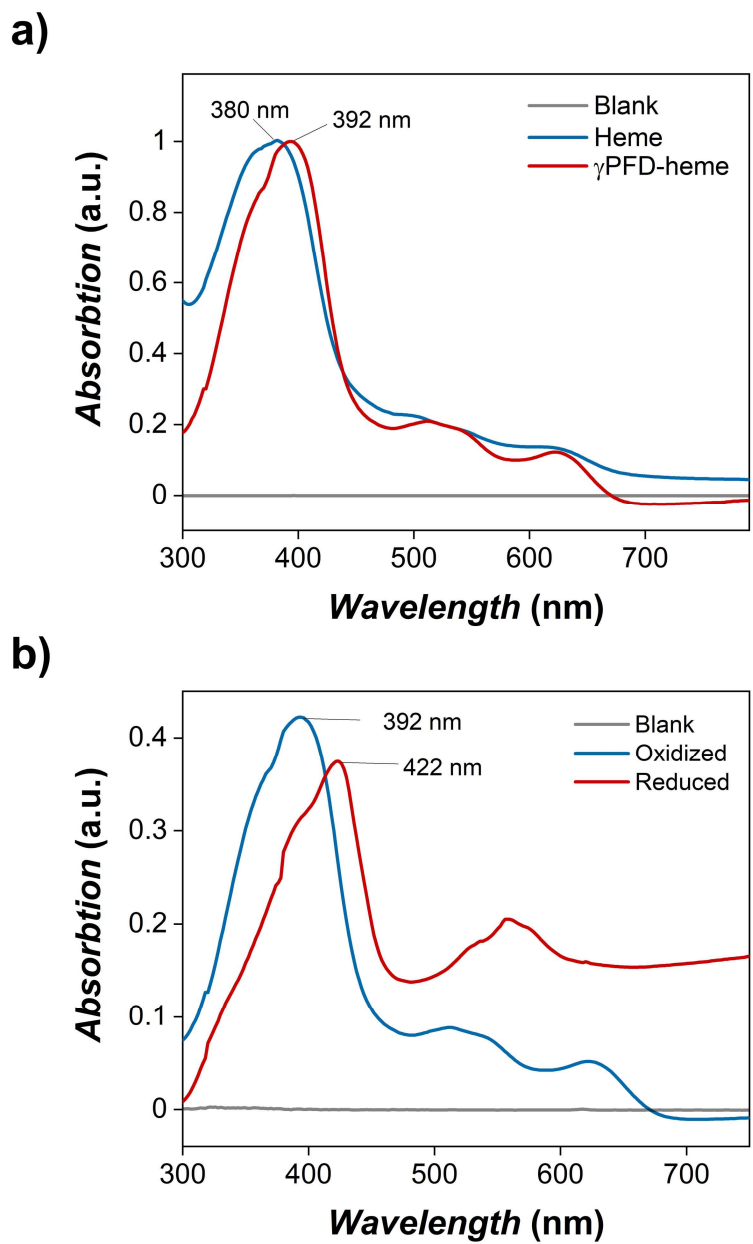

**Figure S1.** Absorption spectra of heme and γPFD-heme nanowires in varying oxidation states. a) Absorption spectra of heme (blue line) and γPFD-heme (red line) in solution that have absorption peaks of 380 nm and 392 nm, respectively. The blank consisted of buffer alone. b) Absorption spectra of a γPFD-heme nanowires in an air-oxidized (blue line) and reduced state (red line) through the addition of sodium borohydride.

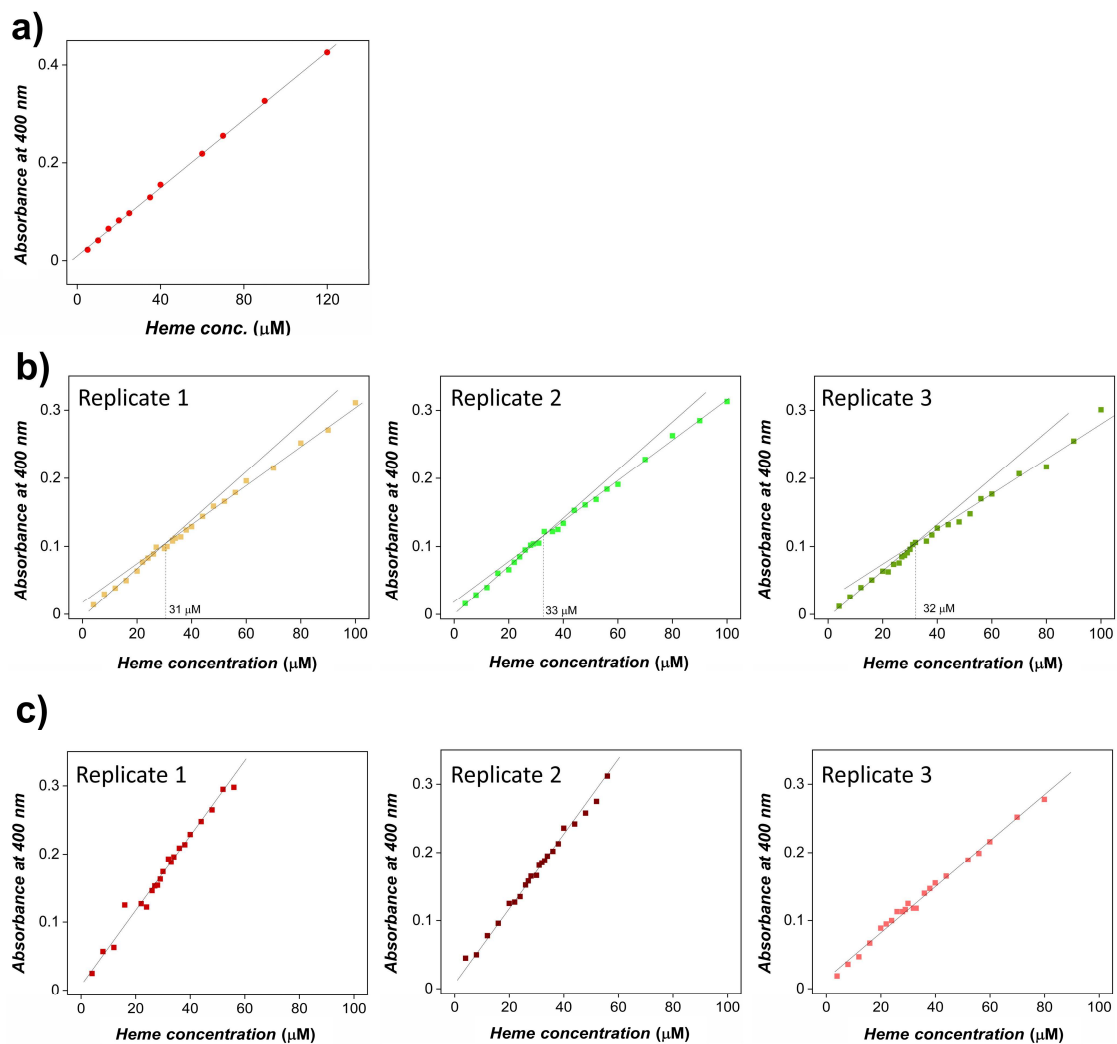

**Figure S2.** Protein-ligand binding isotherms. a) Absorption intensity at 400 nm of heme solution at different concentrations in 150 mM NaCl, 50 mM sodium phosphate, pH 7.4. Binding isotherms with 30 μM protein with varying amounts of heme for b) wild-type γPFD filaments, and c) filaments of truncated γPFD6. Three separate binding experiments are shown for each protein variant.

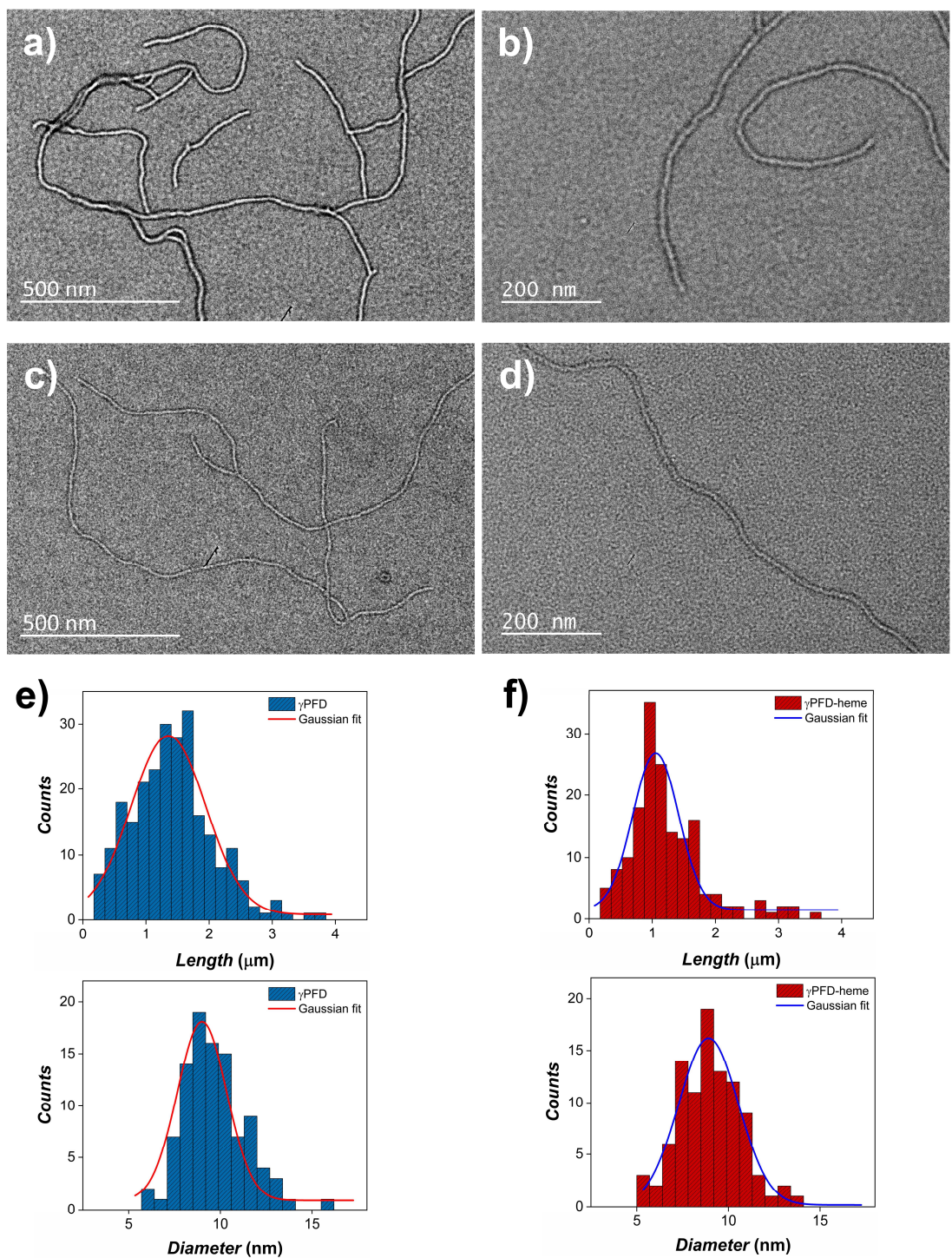

**Figure S3.** Transmission electron microscope images and dimensional analysis of unmodified γPFD filaments and γPFD-heme nanowires. TEM micrograph images of γPFD filaments at a) 10000x magnification and b) 17000x magnification, and images of γPFD-heme nanowires at c) 10000x magnification and e) 17000x magnification. The length and diameter of the filaments and nanowires in the TEM images were measured and plotted as distributions for e) 250 filaments of γPFD and f) 150 γPFD-heme nanowires.

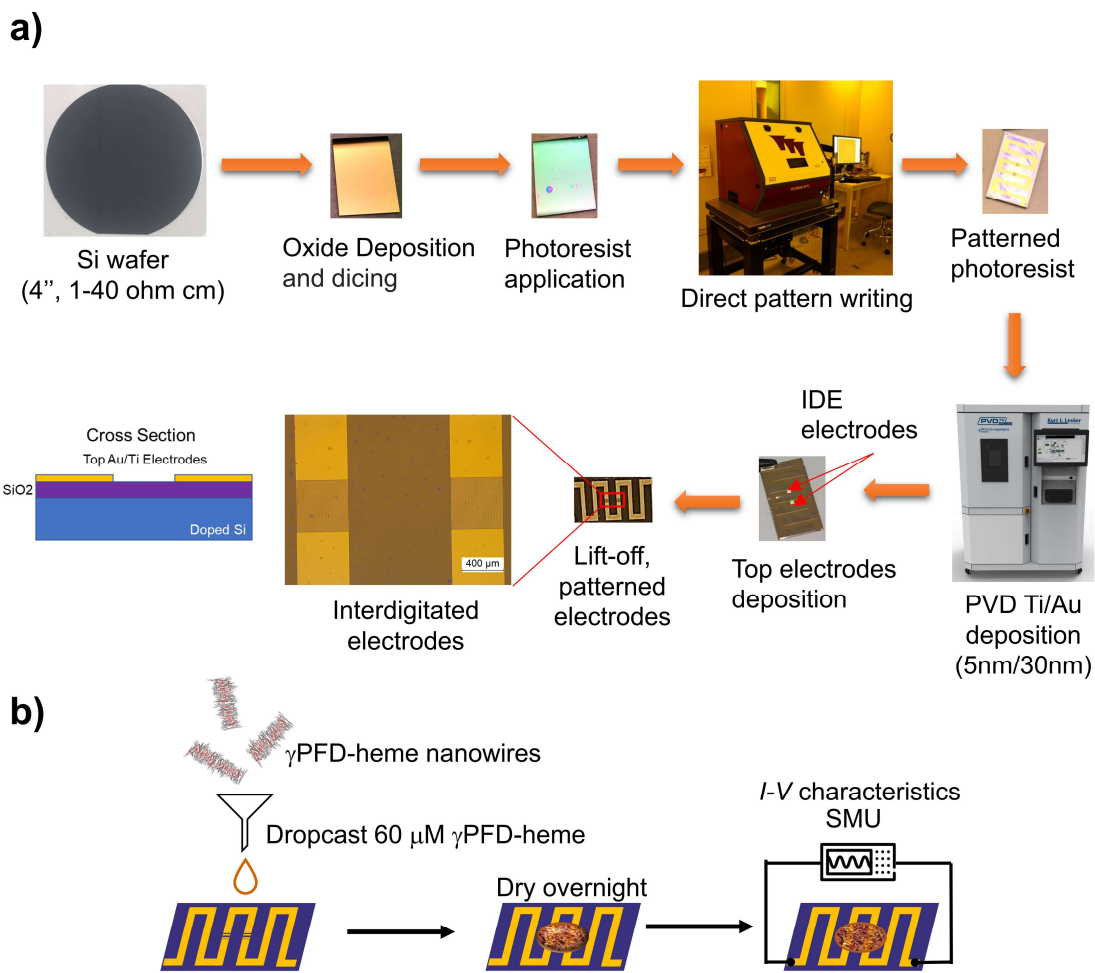

**Figure S4.** Schematic illustration of the fabrication process of an interdigitated microelectrode (IDME) and deposition of γPFD-heme films on IDMEs. a) n-doped silicon wafers (4') were coated with a 250 nm thick $SiO_2$ layer with thermal growth methods consisting of 10% dry oxidation, 80% wet oxidation, and 10% dry oxidation. The wafers were cut into 10 mm x 6 mm chips and the polished side coated with photoresist and hard-baked. The photoresist was patterned using a DMO ML3 Pro direct writer and subsequently developed using RU2020 photoresist developer. A 5 nm titanium layer was deposited on the chip, followed by a 30 nm gold layer using a Lesker E-Beam evaporator PVD75. Finally, the photoresist was removed with acetone to create interdigitated electrodes. b) Schematic illustration of γPFD-heme film deposition on IDME for subsequent analysis by a source measure unit (SMU).

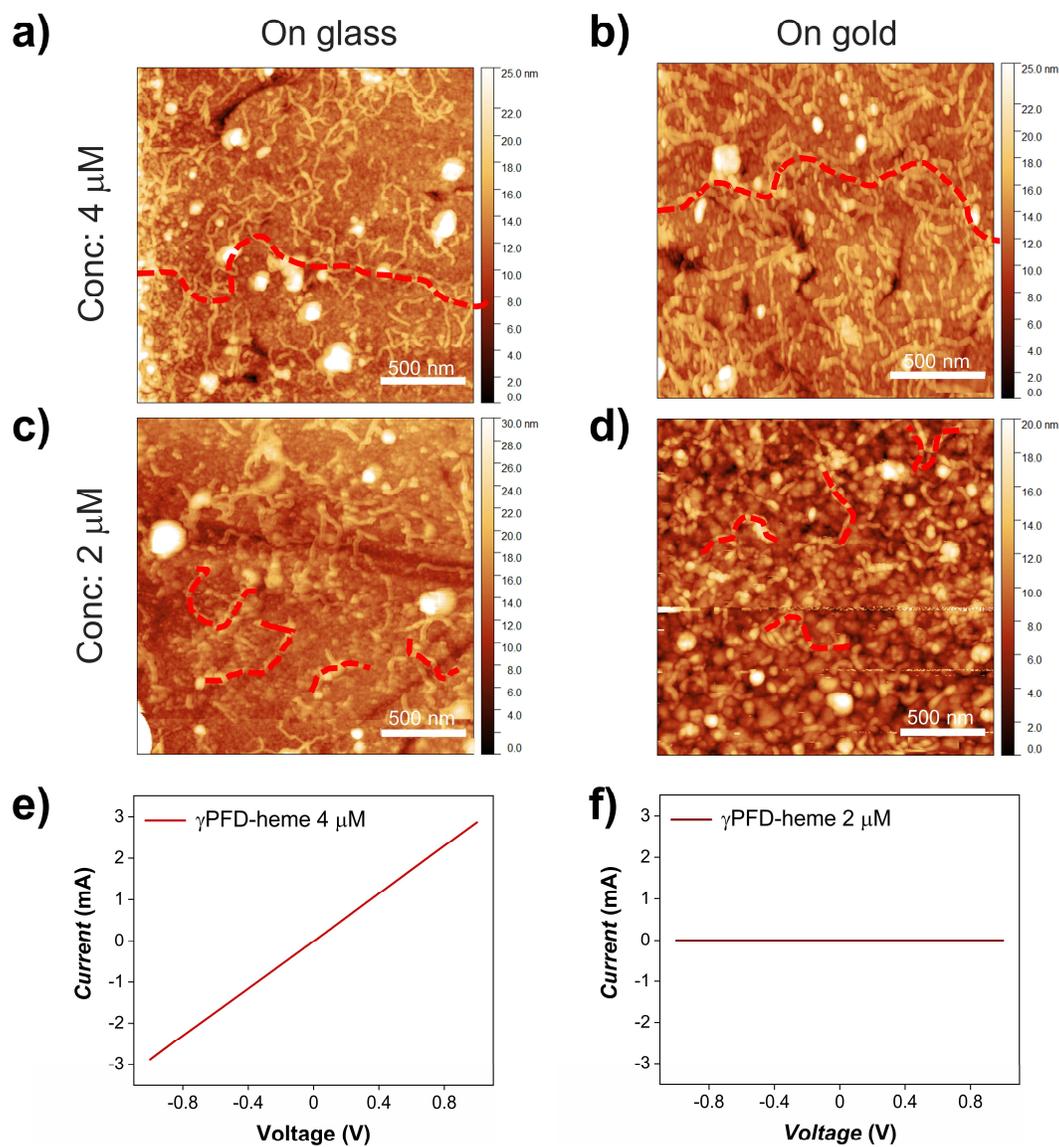

**Figure S5.** AFM analysis of γPFD-heme nanowire morphology on IDME with varying concentrations and relative *I-V* characteristics. A 4 μM solution of γPFD-heme deposited on either the a) glass and b) gold surface shows a continuous path between electrodes marked highlighted in red. A 2 μM γPFD-heme solution deposited on c) glass or d) gold shows a discontinuous path of nanowires highlighted in red. e) *I-V* measurement of the 4 μM γPFD-heme nanowire solution deposited on an IDME and f) *I-V* measurement of the 2 μM solution deposited on an IDME.

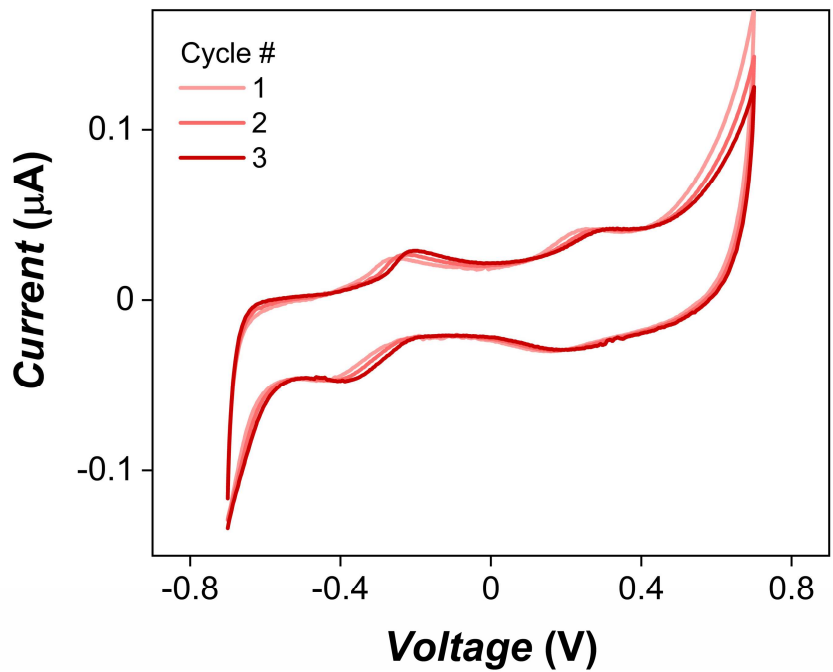

**Figure S6.** Cyclic voltammogram of γPFD-heme nanowires in anaerobic conditions. Cyclic voltammogram cycles of γPFD-heme nanowires performed under anaerobic condition that shows two oxidation and reduction peaks. The two oxidation and two reduction peaks are associated with the anodic and the cathodic voltammetric peaks for each redox process related to $Fe^{3+}/Fe^{2+}$ and $Fe^{2+}/Fe^{+}$ redox of the incorporated heme molecules.

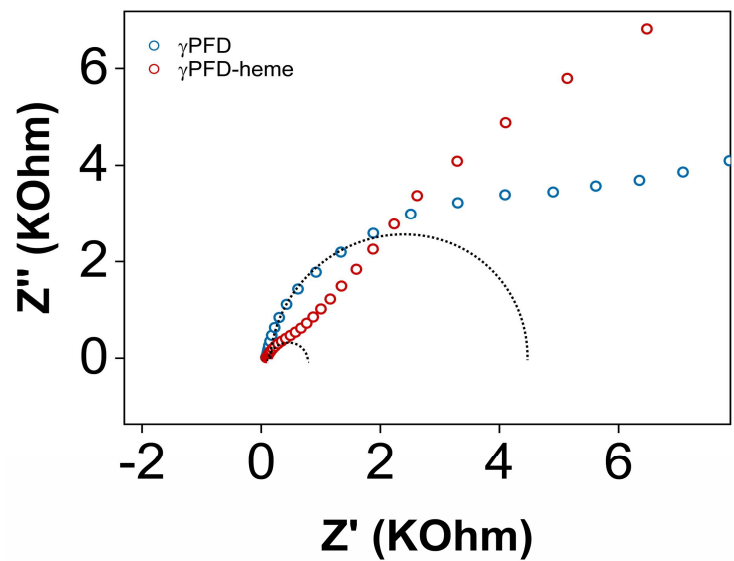

**Figure S7.** Electrochemical impedance analysis of films composed of unmodified γPFD filaments and γPFD-heme nanowires. Nyquist plot at higher frequencies of unmodified γPFD filaments (in blue) and γPFD-heme nanowires (in red).

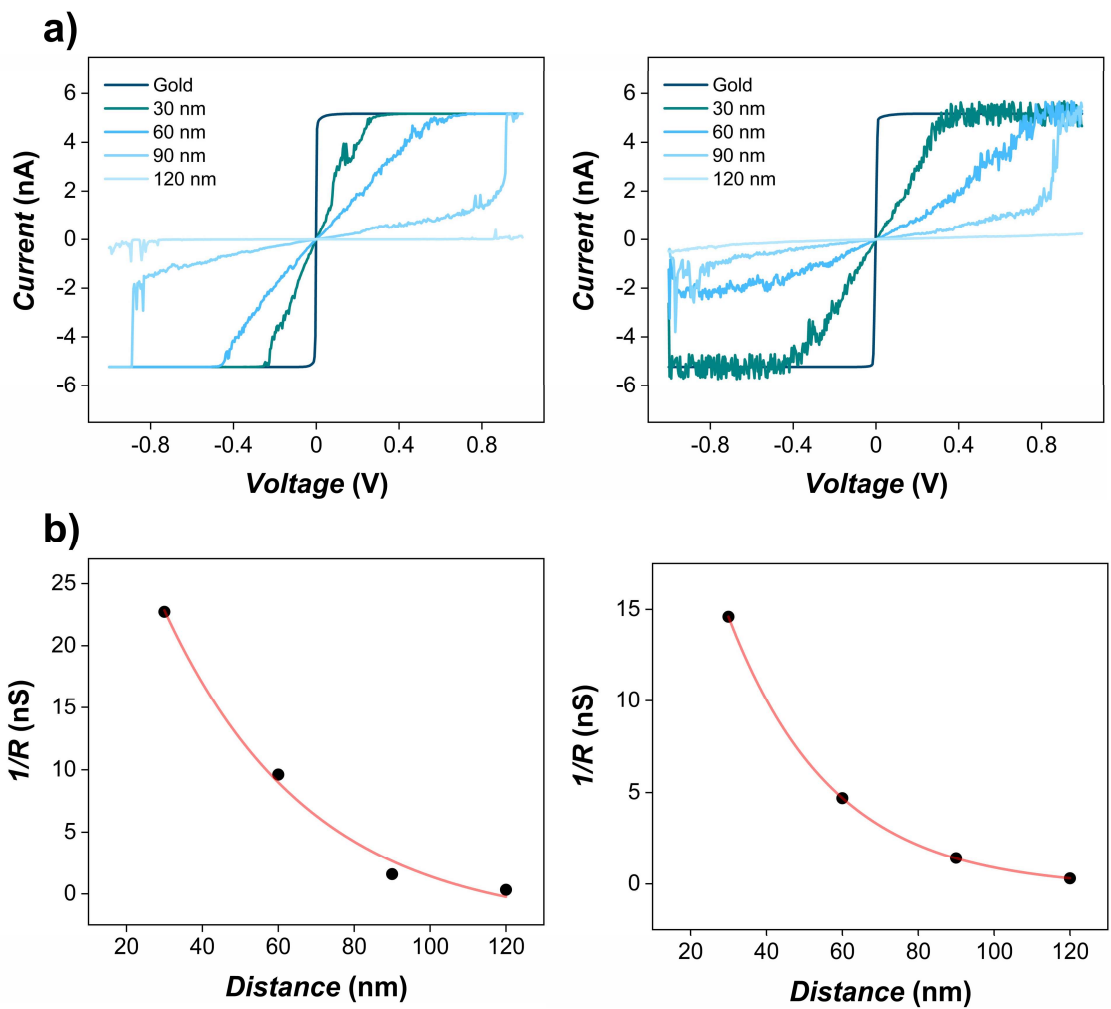

**Figure S8.** Characterization of conductivity and distance dependence of individual γPFD-heme nanowires using c-AFM. a) *I-V* characteristics measured along two different individual γPFD-heme nanowires (left and right panels, respectively) using the c-AFM technique at increasing distances from the gold electrode. b) Conductivity of the individual γPFD-heme nanowire (left and right panels, respectively) in a) were plotted versus the distance from the gold electrode.

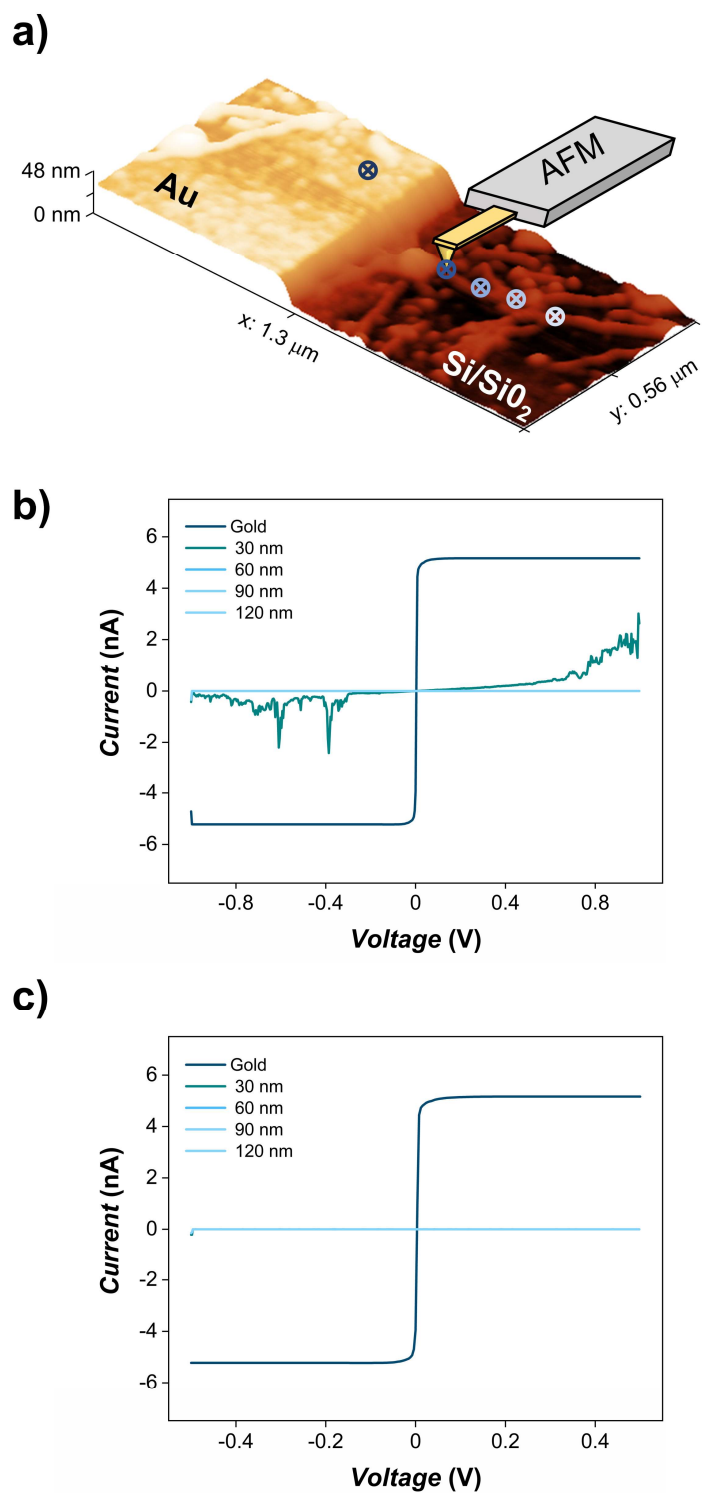

**Figure S9.** Electrical characterization of unmodified γPFD filaments using conductive atomic force microscopy (c-AFM) technique. a) AFM image and scheme of *I-V* measurements performed along the 3D morphology of a γPFD filament on the surface of a gold electrode and Si/SiO$_2$ substrate. b-c) *I-V* characteristics measured along two different individual γPFD filaments using the c-AFM techniques at increasing distances from the gold electrode.

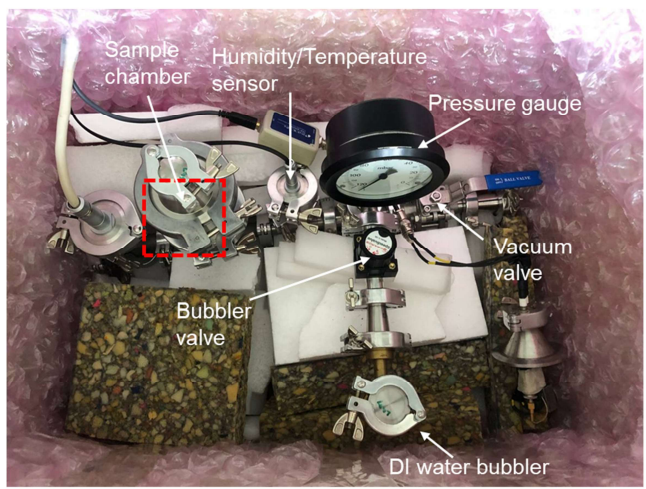 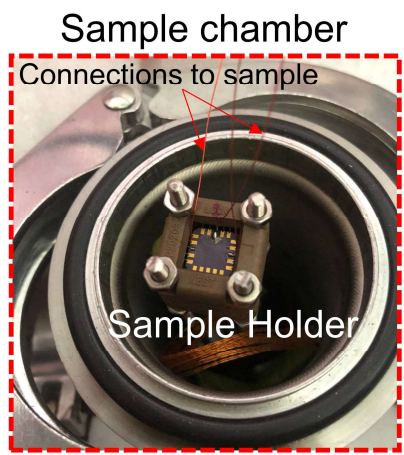

**Figure S10.** Experimental setup for electrical characterization of γPFD-heme nanowire film under controlled relative humidity. Optical images of the humidity chamber setup and sample chamber used to perform electrical characterization of the γPFD-heme nanowire film under controlled relative humidity (RH). The vacuum chamber was designed with a deionized (DI) water bubbler and a valve to regulate the RH inside the chamber. The RH was monitored in real-time through the inclusion of a Sensirion SHT2x capacitive humidity sensor in the sample chamber.

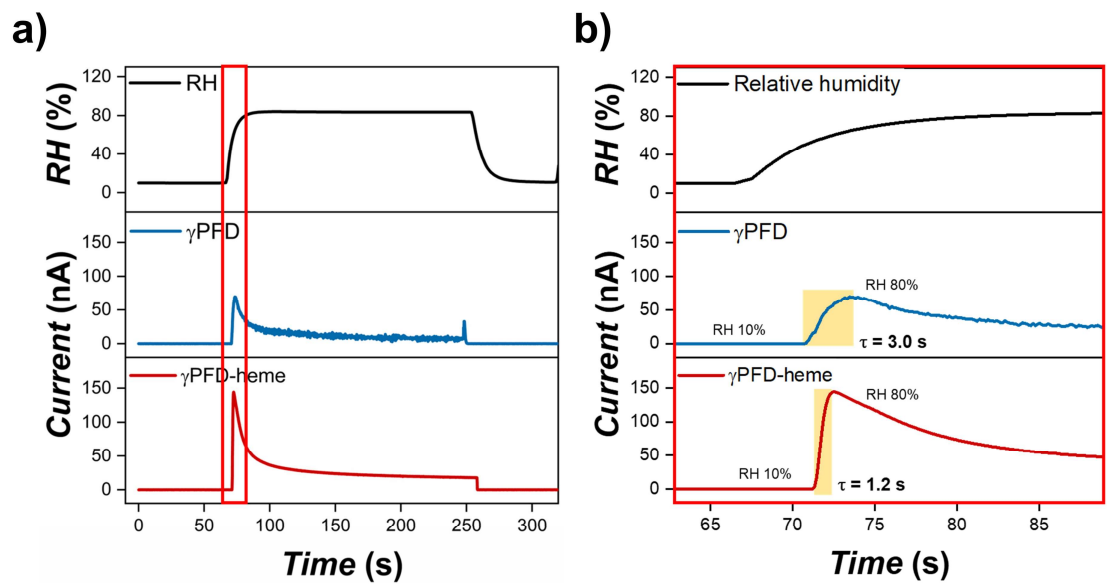

**Figure S11.** Humidity response for a γPFD-heme nanowire-based humidity harvesting device. a) Device output current and b) the expanded section of the initial step highlighted in red, showcasing the estimated time response (τ) after a change from 10% to 80% relative humidity (RH).

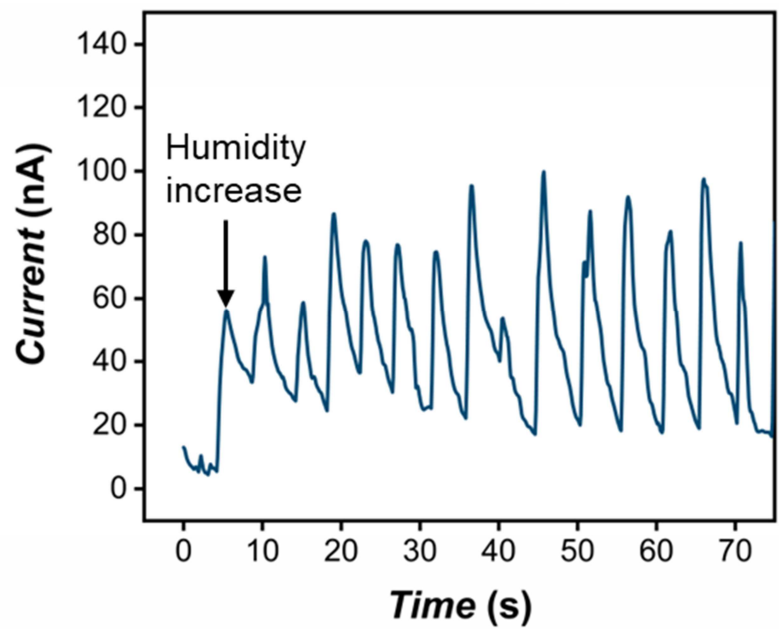

**Figure S12.** Real-time breath monitoring using a γPFD-heme nanowire-based sensor at a breath rate of ~ 3 seconds. The change in current was measured for the γPFD-heme nanowire device while breathing perpendicular on the device surface from a distance of ~ 5 cm, with the application of a constant potential of + 0.3 V between the top and bottom electrode.